\documentclass{JHEP3}
\usepackage{epsfig}

\newcommand{\be}{\begin{equation}}
\newcommand{\ee}{\end{equation}}
\newcommand{\bea}{\begin{eqnarray}}
\newcommand{\eea}{\end{eqnarray}}
\newcommand{\IR}{\mathbb{R}} 
\newcommand{\IN}{\mathbb{N}}
\def\IZ{\relax\ifmmode\hbox{Z\kern-.4em Z}\else{Z\kern-.4em Z}\fi}
\newcommand{\IS}{{\bf S}}

\newcommand{\non}{\nonumber \\}

\def\del{{\partial}}
\def\room{~\rule[-2mm]{0mm}{8mm}}

\def\bw{\bar{w}}

\def\bx{{\bar x}}

\def\al{\alpha} \def\bt{\beta}
  \def\eps{\epsilon}

\def\trho{{\tilde \rho}} \def\hrho{{\hat \rho}}
\def\hmu{{\hat \mu}}

\def\room{~\rule[-2mm]{0mm}{8mm}}

\def\Morse{\mbox{Morse}} \def\merger{\mbox{merger}}
\def\stwosq{{\bf S}^2 \times {\bf S}^2}
\preprint{{\tt hep-th/0206220}}

\title{Topology change in General Relativity,\\
 and the black-hole black-string transition}

\author{Barak Kol \\
 Racah Institute of Physics\\
 Hebrew University \\
 Jerusalem 91904,
 Israel\\
{\tt barak\_kol@phys.huji.ac.il}}

\abstract{In the presence of compact dimensions massive solutions
of General Relativity may take one of several forms including the
black-hole and the black-string, the simplest relevant background
being $\IR^{3+1} \times \IS^1$. It is shown how Morse theory
places constraints on the qualitative features of the phase
diagram, and a minimalistic diagram is suggested which describes
a first order transition whose only stable phases are the uniform
string and the black-hole. The diagram calls for a topology
changing ``merger'' transition in which the black-hole evolves
continuously into an unstable black-string phase. As evidence a
local model for the transition is presented in which the cone
over $\stwosq$ plays a central role. Horizon cusps do not appear
as precursors to black hole merger. A generalization to higher
dimensions finds that whereas the cone has a tachyon function for
$d=5$, its stability depends interestingly on the dimension - it
is unstable for $d<10$, and stable for $d>10$.}

\begin{document}

\begin{flushright} \begin{tabular}{l} {\it To my wife,} \\ {\it Dorit.}
\end{tabular} \end{flushright}

\section{Introduction}

In a gravitational background with compact dimensions several
phases of black object solutions to General Relativity may exist,
some of them having distinct horizon topologies, depending on the
relative size of the horizon (say the Schwarzschild radius) and
relevant length scales in the compact dimensions. As the mass of
the object is varied phase transitions are expected to occur
between the various phases. For concreteness, consider the
simplest case - $3+1$ extended dimensions and a fifth compact
dimension of radius $L$, which is denoted here by the $z$
coordinate. Black objects with Schwarzschild radius much smaller
than $L$ are expected to closely resemble a 5d Schwarzschild
black hole (BH) with a nearly round $\IS^3$ horizon topology,
since by the equivalence principle local measurements at the
vicinity of the BH should not detect the large finite size of the
$z$ axis. Very massive black objects on the other hand, will have
a horizon size much larger than $L$ and will be extended over the
$z$ axis wrapping it completely. Such objects will have $\IS^2
\times \IS^1$ horizon topology and will be referred to as ``black
strings''. One type of black string is the 4d Schwarzschild
solution with the $z$ coordinate added in as a spectator and it
will be called here a ``uniform string'', while another type of
solution may be $z$ dependent, namely, a ``non-uniform string''.

An attempt to build a consistent phase diagram for this system
will lead us to expect a novel topology change within classical
General Relativity. The transition involves a one parameter family
of Ricci-flat metrics which passes through the cone over $\IS^2
\times \IS^2$ (locally).
 This cone may be
deformed in one of two ways much like the conifold transition
leading to different topologies. Each deformation keeps one
incontractible sphere, and ``fills-in'' the other, so the
transition may be described as an $\IS^2$ shrinking to zero size
and then a different $\IS^2$ growing up. In addition to the
similarities with the conifold the differences should be pointed
out as well - this transition happens in any dimension higher or
equal to 5, and there are no Killing spinors.

Gregory and Laflamme (GL) \cite{GL1,GL2} found that the uniform
black string solution develops a perturbative instability at a
certain critical radius, which they interpreted as a transition to
the black hole. Horowitz and Maeda  \cite{HorowitzMaeda} recently
surprised many by arguing that the endpoint of the decay of the
string could not be the black hole, definitely not in finite
affine parameter, basically because of the ``no tear'' property of
horizons. Later Gubser \cite{Gubser} showed that the transition is
actually first order by studying local properties of the
Gregory-Laflamme point.

These papers studied the transition in one direction, namely
starting with the black string and reducing its mass, which will
be denoted here by ``string $\to$ BH'' (not implying that the end
point is necessarily a BH). One may consider also the opposite
direction - starting with a 5d BH and increasing its mass - this
direction will be denoted by ``BH $\to$ string''. Regarding this
latter direction, Susskind \cite{SusskindGW} claimed (based on
Matrix M theory, analogy and intuition) that the transition from
a black hole to a non-uniform string (different from the GL
point) would be analogous to a Gross-Witten transition in 2d
gauge theory, a third order transition (a transition so smooth
that only the third derivative of the free energy jumps).

The original goal of this research was be to determine the phase
diagram for this transition, and especially take the ``BH $\to$
string'' route and identify the critical point where the black
hole develops a perturbative instability, expecting it to be
related to the self intersection of the horizon - the merger of
the north and south poles across the compact dimension.

The motivations were
\begin{itemize}
\item A case study of a rich GR phase diagram.

\item A system with a horizon merger, without time dependence,
which could be a model (both analytically and numerically) for the
heavily studied problem of black hole collision.

In particular, numerical simulations in 4d show (see the review
\cite{Lehner_review} and references therein) that when two BH's
approach each other a cusp forms on the horizon of each BH facing
the other one. Would cusps be present here?

\item A test for the controversy raised by \cite{HorowitzMaeda}.

One would like to determine whether there exists a third
non-uniform phase as claimed by \cite{HorowitzMaeda}, or features
of Gross-Witten transition as claimed in \cite{SusskindGW}.

\item A challenge for a quantum interpretation of a topology change in the horizon.

In string theory or any other theory of quantum gravity a black
hole is a high entropy quantum system (for example, the
black-hole D-brane model). The process of black hole merger
should translate to increased coupling between the respective
quantum systems. It would be interesting to understand the
irreversible merger point in the quantum theory together with the
continuous evolution of the entropy as the two BH's approach each
other, but it will not be discussed in this work.

\item Wide relevance for higher dimensional branes and extra
dimensions.

In string theory branes and particles in low dimension often can
be interpreted as some fundamental branes in 10d or 11d, and then
the issue arises whether they should be considered localized or
smeared over the compact directions, which is exactly the
transition studied here. This system shows up in many other
contexts in string theory as well.

\end{itemize}

Other contributions include the realization of the connection
between the GL instability and negative specific heat
\cite{GubserMitra2,Reall}, a discussion of the charged non-uniform
black string \cite{HorowitzMaeda2}, an attempt to use the Petrov
classification to derive the metric of the non-uniform string
\cite{Smet:2002fv}, a numerical study of a closely related
problem \cite{Wiseman} and other recent papers
\cite{Horowitz_talk,Harmark:2002tr,Hubeny:2002xn,Kang:2002hx,Marolf:2002np,Geddes:2001ht,Casadio:2000py}.

We start in section \ref{review} by setting up the system and our
notation and reviewing a number of central contributions which we
summarize by an incomplete phase diagram (figure
\ref{phasediag1}), looking like an archeological finding with
missing parts.


In section \ref{phase_diagram} Morse theory is used (after a short
review) to demonstrate important constraints on the phase diagram.
I argue why these constraints require a topology change transition
and disfavor a stable non-uniform phase. Finally I present the
simplest (in my opinion), consistent phase diagram (figure
\ref{phasediag2}).

In order to find evidence for a qualitative picture, more
qualitative tools (beyond the Morse index) are sought. In section
\ref{topology} the topology of the various solutions is analyzed,
and it is found that the suggested phase diagram requires a
topology change -- a continuous path of solutions connecting
different topologies. That leads me to seek a local model of the
topology change within Ricci-flat metrics. Moreover, I conjecture
that the point of topology change is also a phase transition
point. In the rest of the section, the topological requirements of
the local transition are elaborated.

The expected Ricci-flat solutions are sought for in sections
\ref{local}, \ref{generalization} and a final summary and
discussion is presented in section \ref{Discussion}.

{\bf Note added:} This paper is the published version of
\cite{TopChange}. Only minor phrasing changes were made to improve
the text but it was not updated. For updates see
 the review \cite{BKrev}.

\section{An incomplete phase diagram}
\label{review}

Consider a static (no angular momentum) black object in an
$\IR^{3+1} \times \IS^1$ gravitational background, namely extended
$3+1$ space-time with a periodic fifth dimension which will be
denoted by the $z$ coordinate. The system is characterized by 3
dimensionful constants: $L$ the size of the extra dimension $z
\sim z + L$ , $M$ the (4d) mass of the system measured at infinity
of $\IR^3$, and $G_5$ the 5d Newton constant. From these a single
dimensionless parameter can be constructed \be \label{mu_M}
 \mu=(G_5 M)/L^2, \ee
while the 4d effective Newton constant is given by $G_4=G_5 /L$.

The isometries of these solutions are $SO(3) \times U(1)$, where
the $SO(3)$ comes from the spherical symmetry in $\IR^3$ and the
$U(1)$ comes from time independence.\footnote{In the Lorentzian
solutions it is the non-compact version of $U(1)$.} The most
general metric with these isometries is \be \label{general metric}
 ds^2 = -e^{2\, A} dt^2 + ds^2_{(r,z)} + e^{2\, C} d\Omega^2, \ee
which is a general metric in the $(r,z)$ plane together with two
functions on the plane $A=A(r,z),\, C=C(r,z)$. The horizon is a
line determined by $e^{2\, A}=0$  and $d\Omega^2=d \theta^2 +
\sin^2{\theta}\, d \phi^2$. In 4d one can choose a coordinate
system that is especially adapted to cylindrical symmetry
\cite{Weyl1917}, but this works best only in 4d.

At least two phases of solutions can be distinguished - a black
string with an $\IS^2 \times \IS^1$ horizon topology and a 5d
black hole with an $\IS^3$ horizon. The uniform black string
(figure \ref{setup1}) is given by the 4d Schwarzschild metric
with $z$ added as a spectator coordinate \be
 ds^2 = -(1-r_4/r)\, dt^2 + (1-r_4/r)^{-1}\, dr^2 + dz^2 +
 r^2\, d\Omega^2 \ee
 where the Schwarzschild radius is $r_4=2\, G_4\, M$.
\EPSFIGURE{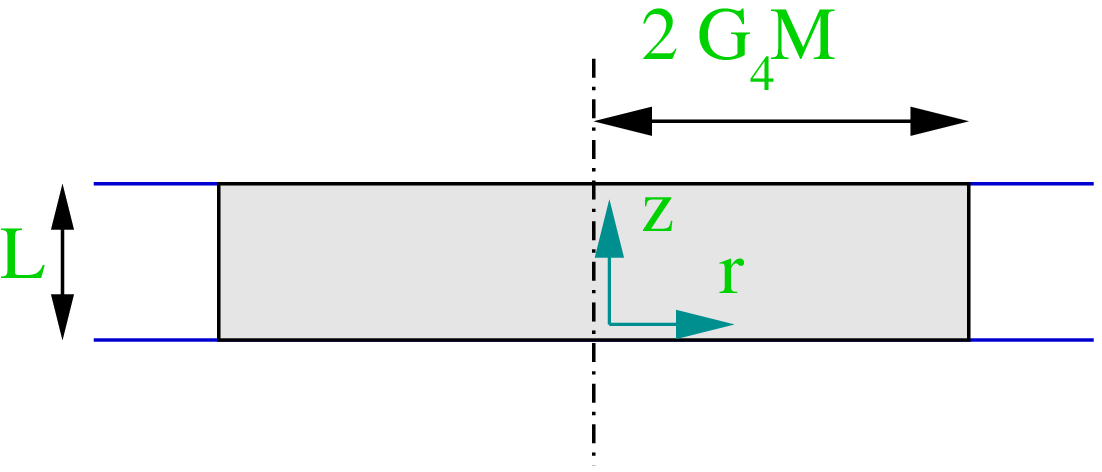}{The uniform black-string. $r$ is the radial
coordinate in $\IR^3$. \label{setup1}}

The small 5d black hole (see figure \ref{setup2})($r_5=\left(
(8/3\pi)\, G_5\, M \right)^{1/2} \ll L$) can be approximated by a
combination of two solutions. Denoting the distance from the
black hole by $\rho$, for $\rho \ll r_5$ it can be approximated
by the 5d Schwarzschild BH \be
 ds^2 = -(1-r_5^{~2}/\rho^2)\, dt^2 + (1-r_5^{~2}/\rho^2)^{-1}\, d \rho^2
 + \rho^2\, d\Omega^2_{~\IS^3} \ee
where $d\Omega^2_{~\IS^3}=d \chi^2 + \sin^2{\chi}\, d\Omega^2$.
For $\rho \gg r_5$ the Newtonian approximation is valid, and the
potential is proportional to the Green function \be
 V(w)={1 \over 4\, r} [\coth (w/2)+ \coth(\bw/2)] \label{5dNewtonian_potential} .\ee
 where $w:=2\, \pi \, (r+iz)/L$.
Indeed close to the source point $w \to 0$, the potential has the
expected 5d behavior $ V \sim (1/4r)[(2/w)+(2/\bw)]=1/(r^2+z^2)$,
while for $\mbox{Re}(w) \to \infty$, the 4d behavior is restored
$\coth(w)=1+2 \exp (-2w), ~ V \sim (1/2r) + O(\exp(-r)$.
 Presumably a full solution can be built perturbatively in
$r_5/L$ from these two approximations. Note that following the
equipotential surfaces of (\ref{5dNewtonian_potential}) one
encounters already a primitive version of a topology change -
close to the source the surface has an $\IS^3$ topology, then
there is one singular surface with a conic singularity (a cone
over $\IS^2$) after which the topology of the surfaces changes to
$\IS^2 \times \IS^1$. \EPSFIGURE{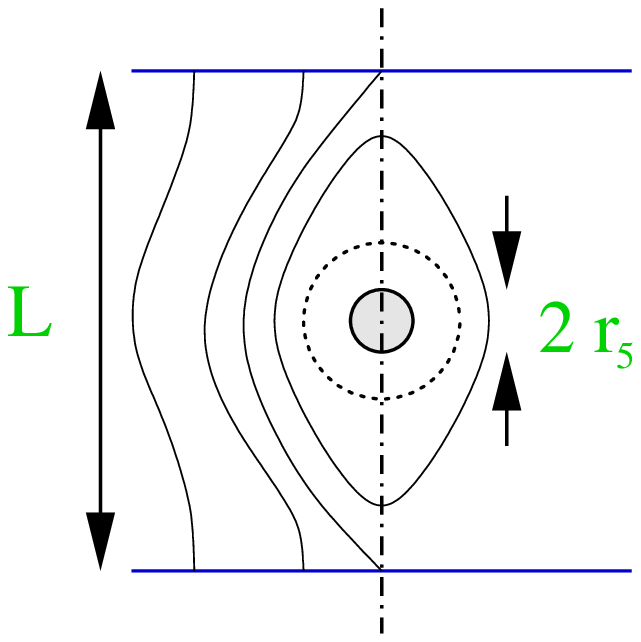}{The 5d black-hole
with Newtonian equipotential surfaces. \label{setup2}}

Comparing the areas of the two solutions one finds that for the
string the area is \be
 A_{string} = 4\, \pi\, r_4^{~2}\, L \sim \mu^2, \label{A_string} \ee
 while for the small BH \be
 A_{BH} \simeq 2\, \pi^2\, r_5^3 \simeq \mu^{3/2}. \label{A_BH} \ee
Hence, for small $\mu$ the black-hole is preferred, while for
larger $\mu$ it is not clear if the black-hole phase exists, but
even if it does than if its area would grow as in (\ref{A_BH}) it
would be dominated by the black-string. Therefore a phase
transition is expected from the outset.

Let us start by assembling the existing knowledge into an
incomplete phase diagram (figure \ref{phasediag1}). The vertical
axis is the parameter $\mu$, while the horizontal axis is an order
parameter\footnote{Even though normally a phase diagram does not
include order parameters.} of non-uniformity $\lambda$. For small
$\lambda$ it is defined by $\lambda:=(1-r_{min}/ \bar{r})$ where
$r_{min},\, \bar{r}$ are the minimum and average respectively of
$r=r(z)$, so that $\lambda=0$ exactly when the solution is a
uniform string. For large $\lambda$ no precise definition is
offered, but consider it to continue to be a measure of
non-uniformity even in the black-hole phase.

\EPSFIGURE{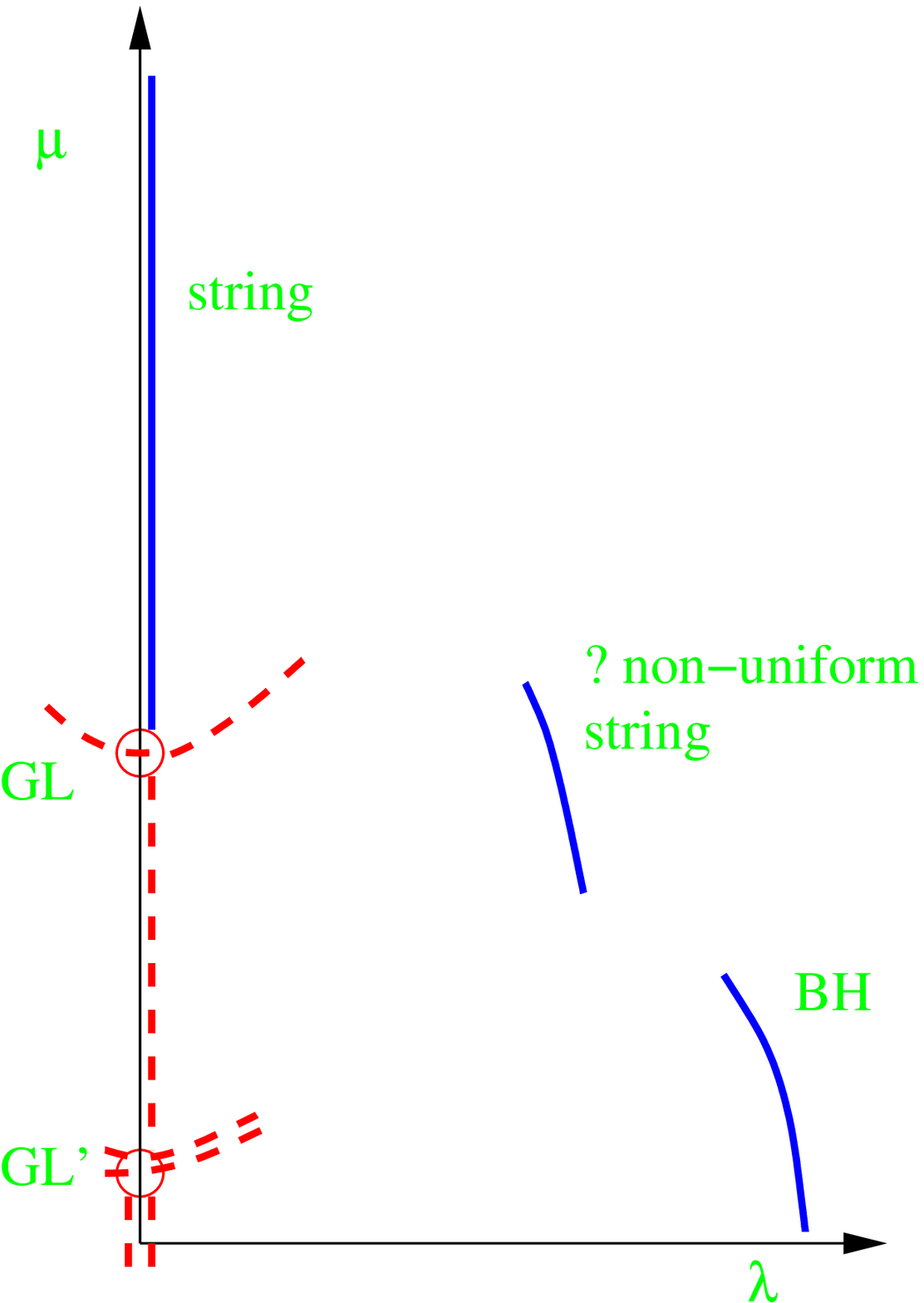,width=10cm}{Collecting present data into
a phase diagram. $\mu$ is the dimensionless parameter of the
problem, and $\lambda$ is an order parameter, a measure of
non-uniformity along the $z$ axis. Stable (unstable) phases are
denoted by a solid (dashed) lines (double dashed lines denote
phases with 2 negative modes). GL denotes the Gregory-Laflamme
critical point at $\mu_{GL}$ and GL' is a copy at $\mu_{GL}/2$.
\label{phasediag1}}

The horizontal axis is inhabited by the uniform string phase,
which changes behavior around the numerically determined
Gregory-Laflamme critical point $\mu_{GL} \simeq .070$
corresponding to $(L/r_4)_{GL} \simeq 7.2$ \cite{GL1}. We denote
the stable (unstable) phase for $\mu > \mu_{GL}$ ($\mu <
\mu_{GL}$) by a solid (dashed) line. The black-hole phase, drawn
to the left, is known to exist for $\mu \ll 1$ but it is not
clear how is evolves for $\mu \sim 1$.

The non-uniform phase (figure \ref{setup3}), argued by Horowitz
and Maeda, is depicted for some intermediate values of $\mu$
where the decaying uniform string could end up. However, we must
pause and be critical here. Since the predicted solution is not
available yet, neither analytically nor numerically
(\cite{Horowitz_talk,unpublished_simulation}),\footnote{Note that
\cite{HorowitzMaeda} came out about a year ago.} we stay on the
side of caution and mark it by a question mark.
\EPSFIGURE{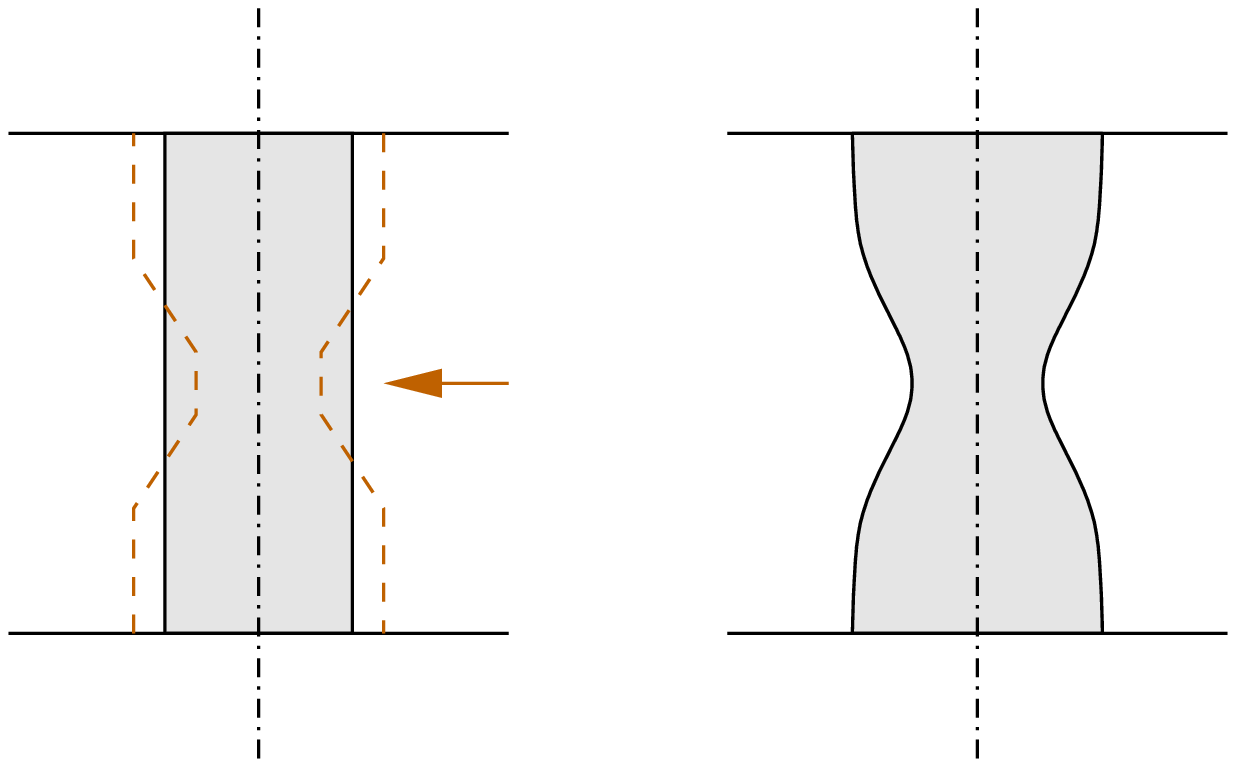}{The Gregory-Laflamme unstable mode, and a
non-uniform black-string \label{setup3}}

At the GL critical point the marginally tachyonic mode leads the
way to a new phase. {\it A priori} this phase could be either
stable or unstable (the two possibilities are shown in figure
\ref{2tran}). In a beautiful work, initiated by the hope of
finding the stable non-uniform phase, Gubser concluded that the
emanating phase is actually unstable \cite{Gubser}. This was done
by following the solution in a perturbative series near the $GL$
point (to third order in some quantities!). Here again we must be
critical, noting that \cite{Gubser} contains a so called
``unwanted scheme dependence'', namely the results depend on an
unphysical parameter (this dependence looks to me like a missing
boundary condition; the local analysis around the GL point is
discussed further in the appendix). However, since the results are
in some sense only weakly uncertain and since the bottom line
agrees so far with unpublished numerical simulations
\cite{Horowitz_talk,unpublished_simulation}, we have some
confidence in it.

It should be mentioned that the $GL$ critical point has infinitely
many copies at $\mu=\mu_{GL}/n$. These are simply a consequence of
the unstable phase of equally spaced $n$ black holes.  Since all
the phases connected to these copies are duplicates of the basic
phases they will be altogether ignored in the rest of the paper.
For example, at $\mu=\mu_{GL}/2$ a string with one unstable mode
turns into a string with 2 unstable modes, and a new phase with 2
negative modes emanates from the point.

The phase transition, like any other one, can be approached from
two different directions, showing possibly different behavior. The
two directions will be denoted by ``string $\to$ BH'' and ``BH
$\to$ string'' but not implying the absence of other phases. So
far we discussed only the string $\to$ BH, but this work will
concentrate on the opposite one.
 \EPSFIGURE{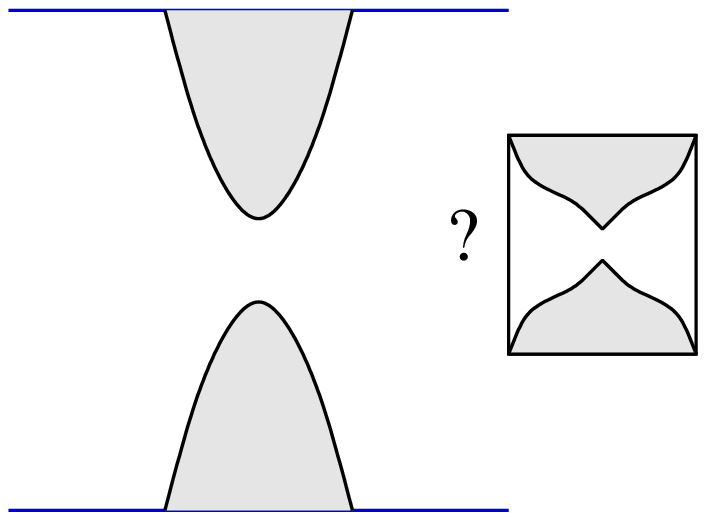}{Approaching merger. Do cusps appear?  \label{setup4}}
\noindent Regarding the latter, BH $\to$ string direction (figure
\ref{setup4}), one should mention the ideas of Susskind
\cite{SusskindGW} on a relation with the Gross-Witten transition
in gauge theory \cite{GrossWitten}, stemming from a matrix theory
description. In the Gross-Witten transition in 2d large $N$ gauge
theories one studies the distribution of eigenvalues of a unitary
matrix, $\exp(i\, \theta)$, the matrix being the Wilson loop over
an elementary plaquette. The behaviour of the eigenvalues is
given precisely by a mechanical model of $N$ point particles
moving on a vertical ring (the bottom of the ring is $\theta=0$)
in the presence of a uniform gravitational acceleration $g$, a
repulsive potential logarithmic at small separations due to the
Van-Der-Monde determinant and a temperature $T$ (see figure
\ref{GW}). It is found that one can take a large $N$ limit in
such a way that the normalized eigenvalue density distribution
approaches a limit $\rho(\theta)$ and a single dimensionless
parameter remains, which is essentially $T$. Moreover, whereas
the probability density at the top $\rho(\pi)$ never vanishes for
finite $N$, it does vanish in the large $N$ limit for small
enough $T$. On the other hand, for very large $T$, the
distribution becomes uniform $\rho(\theta)=1/(2\, \pi)$. The
phase transition occurs at the critical temperature above which
$\rho(\pi)>0$ (see figure \ref{GW}), and it is found to be third
order (so smooth that only the third derivative of the free energy
jumps).

\EPSFIGURE{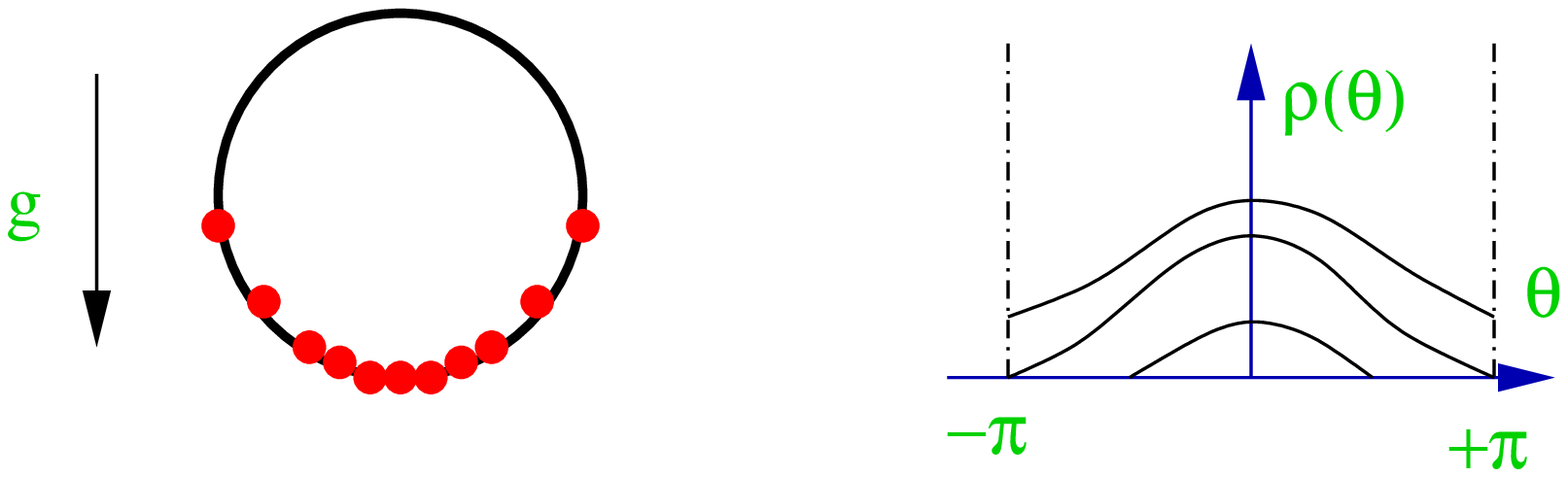,width=14cm}{The behavior of the gas of
eigenvalues in the Gross-Witten phase transition. The three
different density curves are the low temperature ``blob'' phase
analogous to the black-hole, the critical curve, and the high
temperature ``winding'' phase analogous to the black string.
\label{GW}}

According to \cite{SusskindGW}, the distribution of eigenvalues
along the $\theta$ circle is analogous with the mass distribution
of the black hole along the $z$ circle, the low $T$ ``blob'' of
eigenvalues is analogous with the black hole phase, and the high
$T$ ``winding'' phase is analogous with the (non-uniform) string.
\cite{SusskindGW} goes on to predict the existence of a
Gross-Witten like transition in 4d large $N$ gauge theory, based
on the behavior of black holes in a background with a compact
3-torus.

\section{Phase diagram - novel constraints}
\label{phase_diagram}

\subsection{Euclidean version and Free energy}

For Concreteness and convenience, we choose to use the Euclidean
signature for the geometries, and to do thermodynamics with the
free energy, namely in the canonical ensemble. Since the geometry
is time independent we are free to use the Euclidean signature
over the Lorentzian. In the Euclidean version ``time'' is
periodic and the circle fiber degenerates over the horizon, so
the geometry cannot be continued inside the horizon. Another
difference is that Euclidean black hole solutions have a negative
mode not present in the Lorentzian version \cite{GPY} (which is
actually the same as the GL mode, and its eigenvalue gives
$\mu_{GL}$). Hence we make the convention that the number of
negative modes indicated in the phase diagrams is 1 less than
that number for the Euclidean version, to conform with the
Lorentzian picture. From here on we avoid the Lorentzian signature
and any time dependence until the discussion section.

The canonical ensemble offers a direct expression for the free
energy $F=F(\beta)$ through the action $I$ \be
 - \beta\, F = I = {1 \over 16\, \pi\, G} \int_{\cal M} R + {1 \over 8\, \pi\,
 G} \int_{\del {\cal M}} (K-K_0) \ee
where the first integral is the bulk contribution of the Ricci
scalar and the second is a boundary contribution where $K$ is the
trace of the second fundamental form on the boundary, and $K_0$
is the same quantity for a reference geometry which is flat
$\IR^3 \times \IS^1 \times \IS^1$ in our case. When passing to
the canonical ensemble $\beta$ replaces $M$ as a fundamental
variable and so $\mu$ is redefined by \be \label{mu_beta}
 \mu = {\beta \over 8\, \pi\, L} \ee
where the constant factor was determined so that the new and old
definitions would coincide for the uniform string. No confusion
should arise due to the two definitions of $\mu$
(\ref{mu_M},\ref{mu_beta}) since they coincide for the uniform
string which was the only place we used $\mu$ so far, and from
now only $\mu=\mu(\beta)$ (\ref{mu_beta}) will be used.

In our case, the free energy should be thought to be an even
function of $\lambda$. The transformation $\lambda \to - \lambda$
is a symmetry of the action, a remnant of the symmetry of
translations in the $z$ direction, which was fixed by the
symmetry breaking due to the GL mode.

\subsection{Morse theory}

Since our goal of understanding the structure of the phase diagram
is qualitative rather than quantitative \footnote{Namely, we are
interested in the topology of the curve of solutions, and their
stability, rather than its precise location in configuration
space.} we should seek appropriate qualitative tools. As we follow
a curve of solutions, namely an extremum of the free energy, we
may encounter a change in stability or a bifurcation point. These
transitions and other topological properties of extremal points
are the subject of Morse theory, which is our first tool.

We start by reviewing some features of Morse theory. Consider a
function $F: \IR^n \to \IR$. For each extremal point $p$ where
$\vec{\nabla} F=0$  critical groups and a Morse index are defined.
Denoting $F(p)=r,\, F_y=\{ x:\, F(x) \le y \}$, the critical
groups are given by the local relative homology \be
 C_*(F,p) = H_*(F_r,F_r \backslash \{p\}), \ee
where both sets are intersected with a small enough neighborhood
of $p$. The Morse index is the (Euler) index of this homology \be
I_{\Morse}(F,p) = \sum_q (-)^q\, \mbox{dim}(C_q(F,p)). \ee

For the generic case of a non-degenerate Hessian $\del_{ij} F(p)$,
 $n$ the number of its negative eigenvalues, is called
``the Morse type'', the critical groups are $C_n=\IZ$ and all the
others vanish, and the index is \be
 I_{\Morse}=(-)^n \ee
 For a more general extremal point the index is given by
considering the function $\vec{\nabla} F/\,|\vec{\nabla} F|$
defined on a small sphere that surrounds $p$ as function from
$\IS^{n-1}$ to itself and as such it has a winding number $w$ and
one defines \be
 I_{\Morse}=w(\vec{\nabla} F(p)). \label{winding_index} \ee
 One verifies that the previous definition is indeed a special
case of the latter.

One of the basic results is that if we take an arbitrary sphere
$S \subset \IR^n$ with no extremal points then \be
 w(\vec{\nabla} F\,(S)) = \sum_p I_{\Morse}(p) \ee
where the sum extends over all the extremal points inside $S$.

\EPSFIGURE{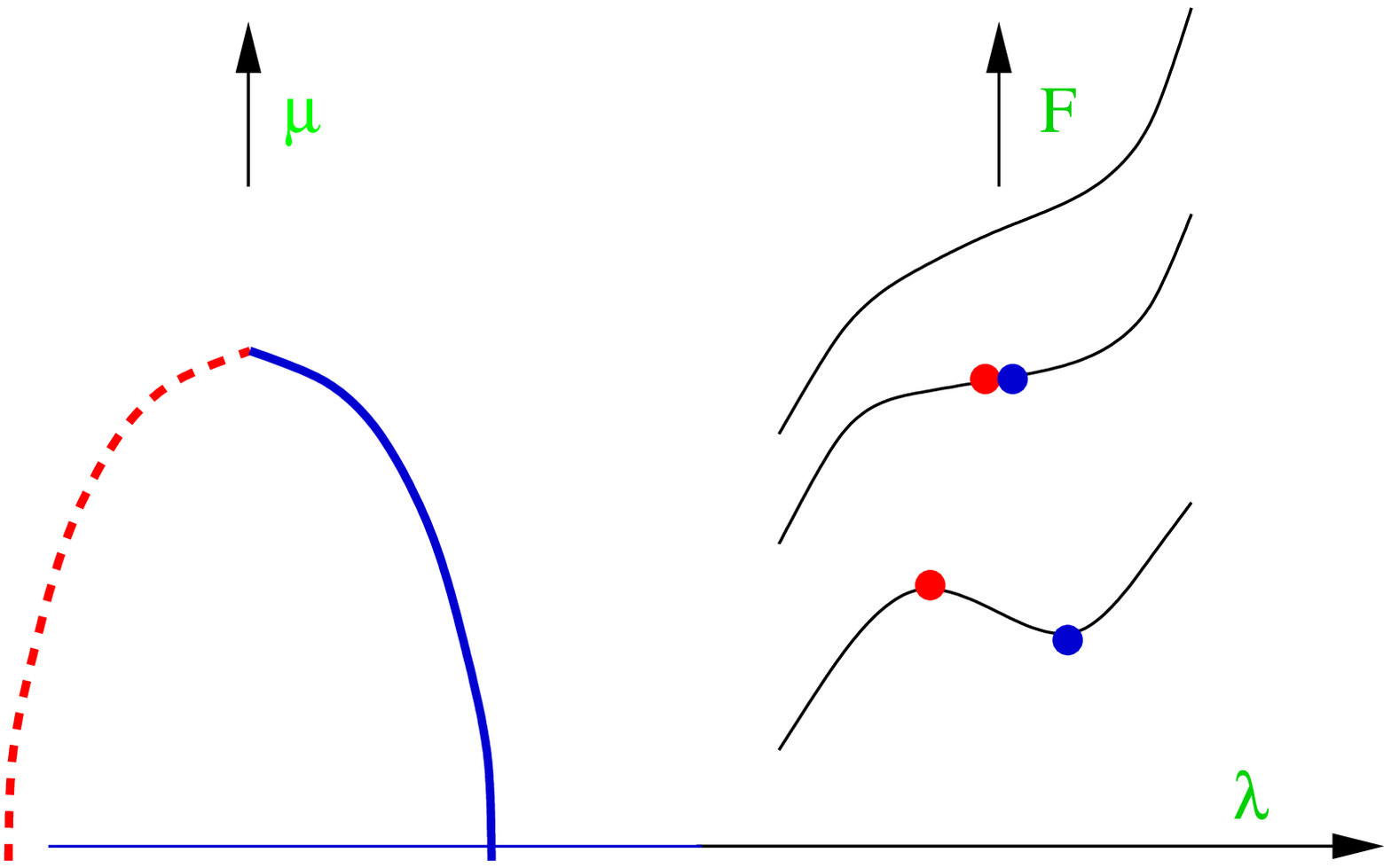,width=7cm}{The most generic phase transition
- first order with a cubic term. Stable (unstable) phases are
denoted by solid (dashed) lines, and an arbitrary number of
spectating negative modes $n$ may be added. \label{1tran}}

As a parameter is changed transitions may occur but the Morse
index (and other quantities) stay invariant. The most generic
transition is the annihilation (or creation) of a $n,n+1$ type
pair. For example in 1d it is the creation of a maximum and
minimum pair, and can be modelled by the equation $F(\lambda)=
\mu\, \lambda + \lambda^3$ (see figure \ref{1tran}). The more
general case can be gotten by adding $n$ spectating negative
modes.

Out of this primitive vertex one can create more elaborate
vertices. The most generic vertex for an even $F$ (figure
\ref{2tran}) can be modelled in 1d by $F(\lambda)= \mu\, \lambda^2
- \lambda^4$. It describes a solution with $n=1$ bifurcating into
2 solutions with $n=1$, and one stable solution (one can think of
this vertex as the primitive vertex glued to a free line with
$n=1$).

\EPSFIGURE{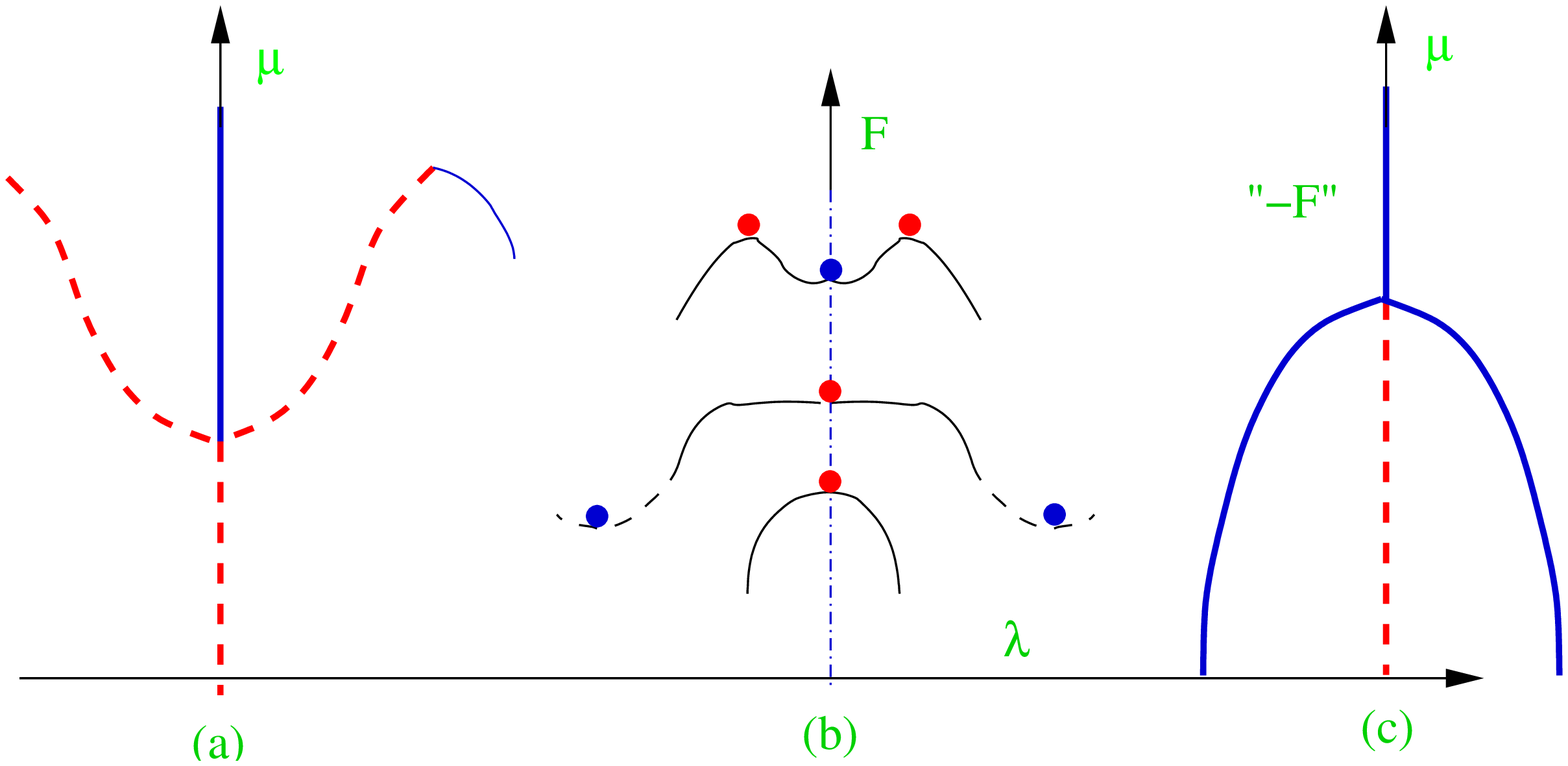,width=14cm}{The generic phase transition with
even free energy. (a) is the phase diagram corresponding to the
free energy $F$ depicted in (b) while (c) is the diagram that
corresponds to $-F$ (and $-\mu$). \label{2tran}}

It can be seen from these examples that it is useful to think
about Morse theory as a {\it conservation rule} for extremal
points under deformations, namely an extremum of type $n$ can only
be ``annihilated'' by a vertex with its ``anti-particle'', an
extremum of type $n+1$.

\subsection{A suggested phase diagram}

Let us apply the previous discussion to find consistent ways to
complete the phase diagram in figure (\ref{phasediag1}). From the
Morse theory ``conservation rule'' it is now apparent that this
diagram has a serious problem -- if one assumes that a stable
non-uniform string phase exists for $\mu \to 0$ \footnote{which is
in spirit of \cite{HorowitzMaeda} as we discuss shortly.} there is
an inconsistent non-conservation of phases between $\mu \to 0$ and
$\mu \to \infty$. The two ``completions'' of figure (\ref{wrong})
exemplify the problem: diagram (a) has an inconsistent vertex
between stable phases, while diagram (b), though consistent,
retains the paradox of Horowitz-Maeda \cite{HorowitzMaeda} intact
since the non-uniform phase will have to decay now into a BH.

\EPSFIGURE{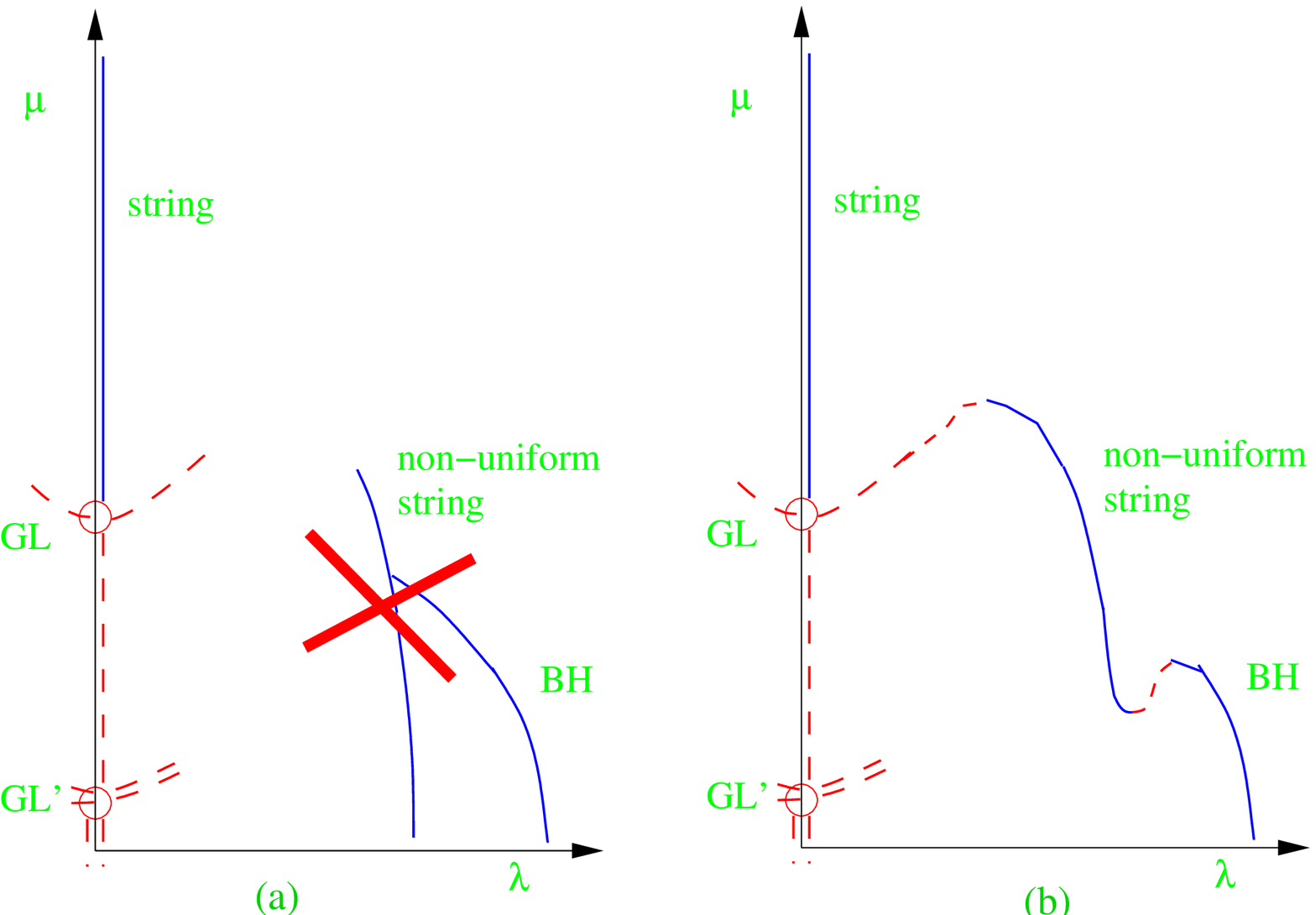,width=13cm}{Disfavoured completions of the
phase diagram. Diagram (a) is inconsistent as it does not respect
the Morse index at the vertex of the black hole and the
non-uniform black string. Diagram (b) is consistent but the
non-uniform string disappears for small $\mu$ and the paradox of
\cite{HorowitzMaeda} moves to the second transition.
\label{wrong}}

In order to solve the problem there must be either a new phase at
the asymptotic regimes $\mu \to 0$ (decompactified 5d) or $\mu \to
\infty$ (4d) or a runaway phase (a phase which runs to infinity in
field space at finite $\mu$) \footnote{All of these options are
problematic, and I speculate here on theorems that could eliminate
them. It is believed (see also \cite{Gubser} section 3.7 and a
reference therein) that for $\mu \to \infty$ (4d) the only
solution is the uniform string. One way (suggested to me by J.
Maldacena) is to show that as $\mu \to \infty$ the regime of
validity of the perturbative analysis around the 4d Schwarzschild
solution grows indefinitely. Another way would be to compute the
index from (\ref{winding_index}). Since the field of metrics is
infinite dimensional a winding number cannot be defined, but as
long as our critical points have finite $n$ it is plausible that
the infinite dimension could be regularized by a UV cutoff. New
localized solutions in decompactified 5d, on the other hand, are
forbidden by a generalization of the uniqueness theorem
\cite{Hwang,Gibbons:2002bh}.}.
 Due to these difficulties, and since the stable non-uniform black-string was not constructed
yet neither analytically nor numerically, I choose to {\it
eliminate it} altogether and suggest the following phase diagram
(figure \ref{phasediag2}), which is the simplest option in my
opinion. In the suggested phase diagram an unstable non-uniform
string emanates from the black string at $\mu_{GL}$ using the even
vertex (\ref{2tran}) and then simply annihilates the black hole
phase, using a generic vertex (\ref{1tran}). The latter transition
will be called a ``merger transition'' and its $\mu$ will be
denoted by $\mu_{\mbox{merger}}$. Here the point of phase
transition is identified with the point of horizon topology
change, but {\it a priori} these two points are independent.
Admittedly, other consistent phase diagrams can be imagined, but
this is the simplest one in my opinion. Therefore the challenge
now is to seek some confirmation, in which process we will find
that changes are due in the neighborhood of the merger transition.

\EPSFIGURE{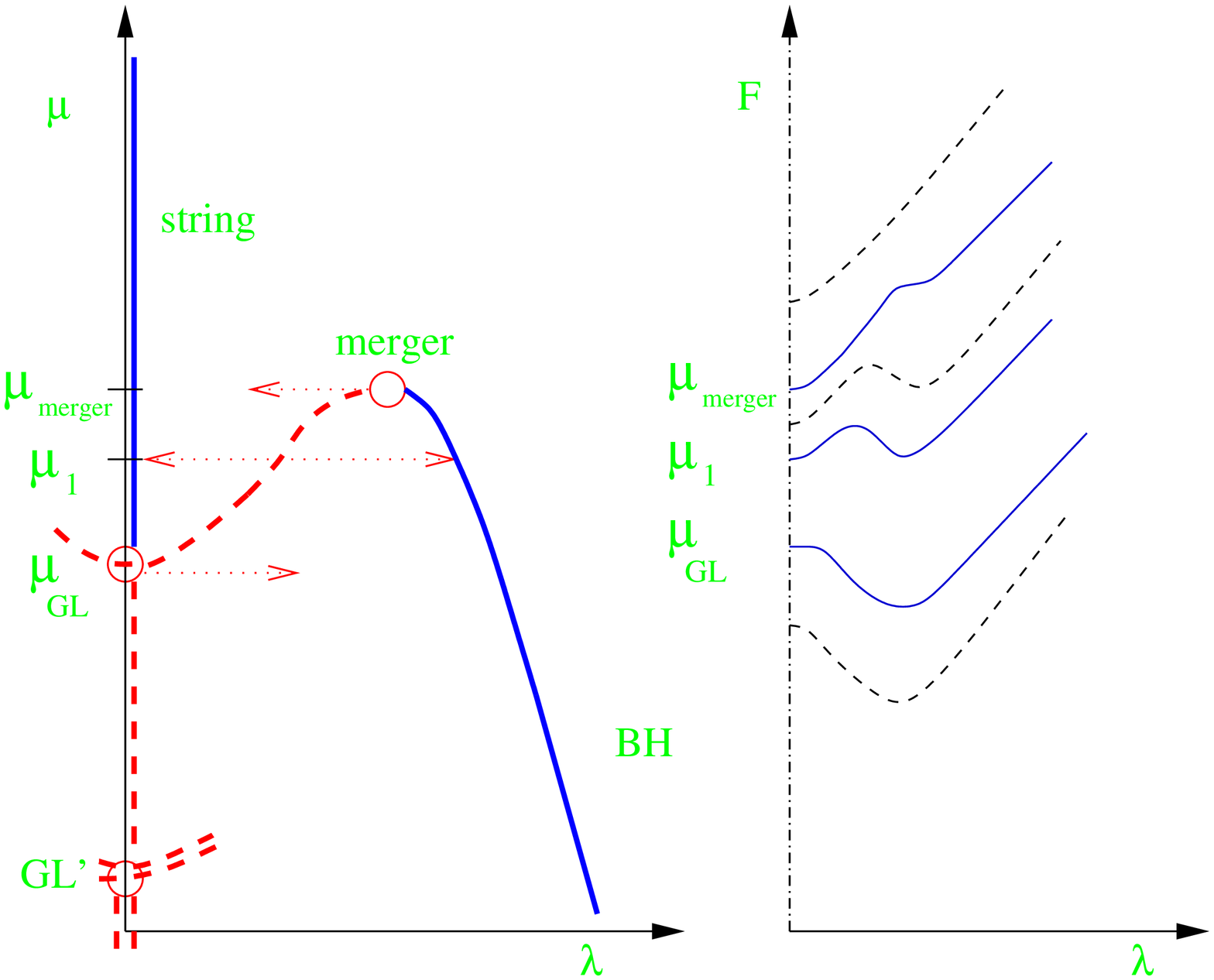,width=15cm}{(Left) A suggested phase
diagram, which will be refined further in the area of the merger
transition. $\mu$ is the dimensionless parameter, and $\lambda$ is
an order parameter (measure of non-uniformity). Dotted lines
denote transitions - a first order transition at $\mu_1$ and
tachyonic decay from the two other points. (Right) Curves of the
free energy are given for several $\mu$ values, solid lines
denoting critical graphs for $F$.
 \label{phasediag2}}

\section{Topology - prediction}
\label{topology}

Let us use another qualitative tool at our disposal namely the
topology of the solutions. The black string is contractible to
$\IS^2 \times \IS^1$ by contracting the $(r,t)$ plane to a point,
hence \be \mbox{string} \simeq \IS^2 \times \IS^1 \ee where the
equivalence denotes homotopic (and homological) equivalence.

The 5d BH is a direct sum of the $\IR^3 \times \IS^1_z \times
\IS^1_t$ background with the 5d Schwarzschild solution which is
topologically a disk bundle over $\IS^3$. This topology has the
following non-contractible cycles - an $\IS^1$ along the $z$
axis, an $\IS^3$ along the horizon, and a less obvious $\IS^2$
(see figure \ref{merger}) which is made out of the segment of the
$z$ axis connecting the north and south pole, together with the
$t$ fiber, which degenerates on the horizon to complete the
sphere.

We see that {\it the two solutions have different topologies}
since the non-contractible 3-cycles have a different topology.
This is a much stronger statement than the change in horizon
topology. Usually one thinks about the configuration space of GR
as being disconnected between different topologies, namely one is
given a manifold on which to solve Einstein's equations together
with some boundary conditions, but the topology of the manifold is
part of the data, rather than the dynamics. On the other hand we
know already about several topology change transitions in 6d or
higher such as the flop and the conifold which occur at finite
distance in moduli space. Therefore we continue by assuming that
such a topology change occurs in our 5d system as well, and seek a
local model for it.

Analyzing the topology near the merger transition (see figure
\ref{merger}) we see that the topology of a 4d boundary away from
the wall (which hence remains constant during the transition) is
$\IS^2_{\theta,\phi} \times \IS^2_{r,z,t}$ - the line in the
$(r,z)$ plane combines with the $t$ circle fiber to give
$\IS^2_{r,z,t}$ over which the $\IS^2_{\theta,\phi}$ is fibered at
each point and the fibration can be seen to be trivial due to the
decoupled form of the metric \ref{general metric}. In the
black-hole phase the $\IS^2_{\theta,\phi}$ shrinks completely,
while in the string phase the $\IS^2_{r,z,t}$ shrinks. So the
topology change can be modelled by the ``pyramid'' familiar from
the conifold transition, and {\it the singular solution is the
cone over} $\IS^2 \times \IS^2$.

\EPSFIGURE{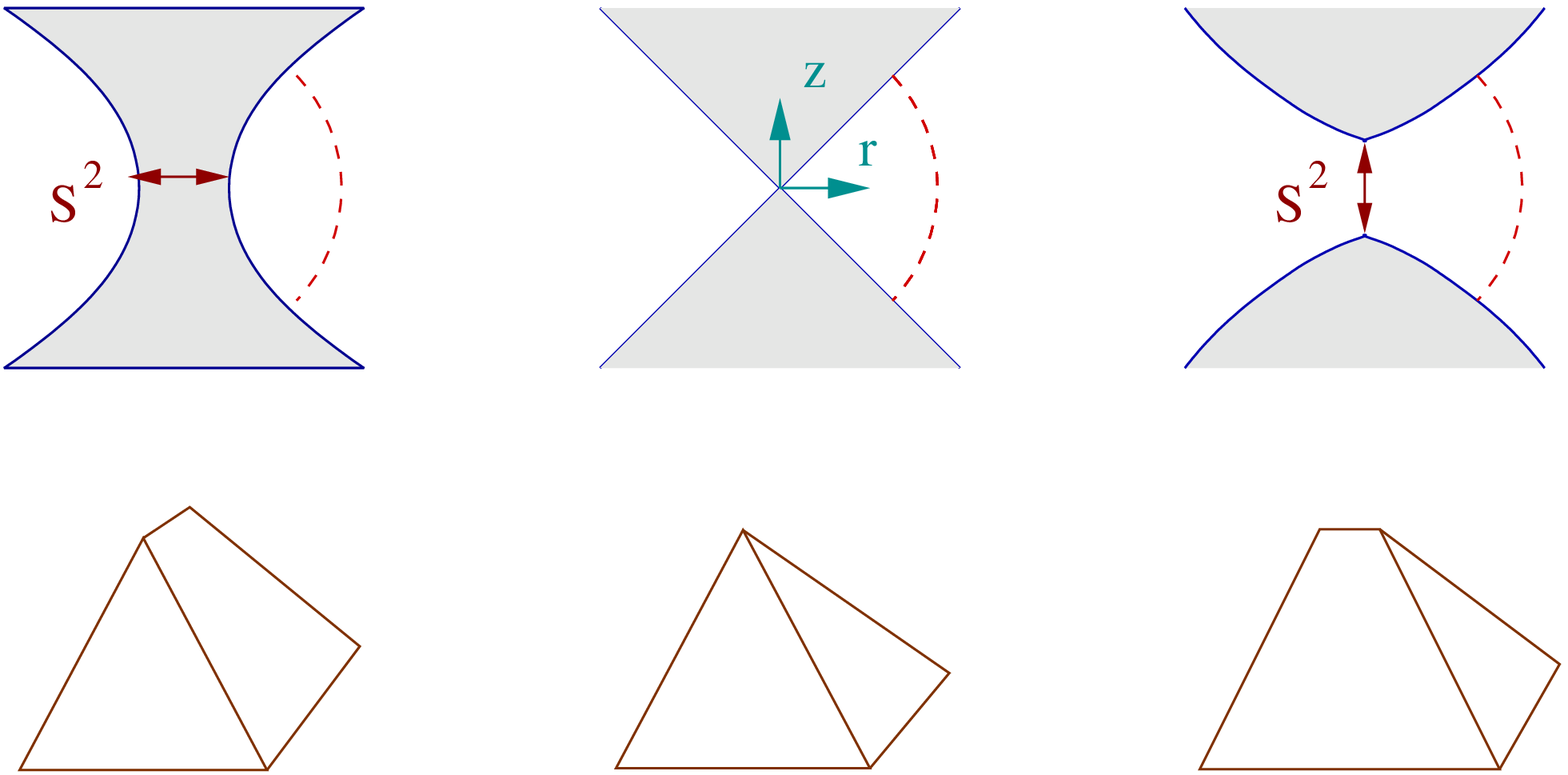,width=14cm}{The merger transition. Shaded
regions are inside the horizon and the dashed line is the base.
The singular configuration is a cone over $\stwosq$.
\label{merger}}

Topologically the transition can be modelled by the equation \be
 |\vec{x}|^2-\epsilon=|\vec{y}|^2 \ee
where $\vec{x} \in \IR^3$ is a 3-vector and , $\vec{y}=(r,t,z)$
combines the $z$ coordinate nicely with $(r,t)$, the origin of $z$
being at the merger point, and the origin of $(r,t)$ on the
horizon. For $\epsilon>0$ we have the string, while for
$\epsilon<0$ it is the black hole. We view {\it the realization of
the transition within the class of Ricci-flat metrics as a
challenge for our picture}.

\section{A local topology change - confirmation}
\label{local}

It is required to find Ricci-flat metrics with $SO(3) \times
U(1)$ isometry which realize the topology change through the cone
over $\IS^2 \times \IS^2$.

\subsection{The cone}

We start by looking for a Ricci-flat metric for the cone itself, $
ds^2=d\rho^2 + \rho^2\, ds^2_{\IS^2 \times \IS^2}$, which is
equivalent to finding an Einstein metric on $\IS^2 \times \IS^2$
with a positive Einstein constant 3 (the unit $\IS^2$ has Einstein
constant 1).
 I find that the most general such metric is $
 3\, ds^2_{\IS^2 \times \IS^2} = d\chi^2 + e^{2c}\, \sin^2(\chi)\, d\sigma^2 + d\theta^2 +
 \sin^2(\theta)\, d\phi^2$, where $c$ is an arbitrary constant which will be set to zero to
avoid a unwanted singularity which persists away from the apex.
So the geometry is just a product of two round $\IS^2$'s with the
angle of the cone chosen to assure Ricci flatness \be
 ds^2 = d\rho^2 + {\rho^2 \over 3}
 \left( d\Omega^2_{\theta,\phi} + d\Omega^2_{\chi,\sigma} \right). \ee
Note that the isometry of the local solution enhances to $SO(3)
\times SO(3)$.

Let us verify that this is the only possibility. The $SO(3) \times
U(1)$ isometry generates homogeneous 3d orbits $[SO(3) \times
U(1)]/U(1)$ which are classified by an integer $k$ giving rise to
$\IS^3/\IZ_k$ -- a circle bundle over $\IS^2$ with monopole charge
$k$ (with the $\IZ_k$ acting along the Hopf fibration). Then
$\IS^3/\IZ_k$ must be fibered over an interval (or circle but that
would give a trivial fibration and the wrong topology) and
``sealed off'' at the end-points to make a compact manifold of
topology $\IS^2 \times \IS^2$. The possible metrics are $ds^2=
d\chi^2 + e^{2\,A}(d\theta^2 + \sin^2(\theta)\, d\phi^2) +
e^{2\,B} (d\sigma -k\, \cos(\theta)\,d\phi)^2$. Moreover the
discrete $\IZ_2$ isometry of our solution for reflection of the
$z$ axis, translates into the functions $A=A(\chi),~B=B(\chi)$
being even. If the function $e^{2\, B}$ has zeroes at the end of
the interval then we get the Hirzebruch spaces, ${\bf F}_k$ -
fibrations of ${\bf CP}^1$ over ${\bf CP}^1$. Therefore the only
way to get the product topology $\stwosq$ is to start with $k=0$.
Then the Einstein equations determine $e^{2\, A}=c_1,\, e^{2\,
B}=c_2\, \sin^2(\chi)$, as claimed in the previous paragraph.

\subsection{The deformed cone}

We now proceed to look for Ricci-flat solutions over the deformed
cone. {\it A priori} I  do not know of any obstruction to do that:
in 2d the deficit angle of the cone would impose some non-zero
curvature (and Ricci) due to Gauss-Bonnet, but that does not apply
here. 

The ansatz is \be \label{def_cone_ansatz}
 ds^2 = d\rho^2 + e^{2\, a} (d\theta^2 +
\sin^2(\theta)\, d\phi^2) + e^{2\, b} (d\chi^2 + \sin^2(\chi)\,
d\sigma^2) \ee
 where $a=a(\rho),~b=b(\rho)$ and we retained the enhanced $SU(2) \times SU(2)$ isometries.

The equations are \bea \label{def_cone_eq}
 a'' &=& e^{-2\, a} -2\, a'^{~2} - 2\, a'\, b' \non
 b'' &=& e^{-2\, b} -2\, b'^{~2} - 2\, a'\, b' \non
  0  &=& a'^{~2} + a'' + b'^{~2} + b''   \eea
One verifies that the 3 equations for 2 unknown functions imply a
first order constraint
 \be  0   = a'^{~2} + 4\, a'\, b' + b'^{~2} - e^{-2\, a} - e^{-2\, b} \ee
which is consistent with the second
order equations, and reduces the number of required initial
conditions from 4 to 3.

The cone solution corresponds to $a_0=b_0=\log(\rho)-\log(3)/2$.
Let us linearize first the equations around this solution.
Denoting \be \begin{array}{lr}
 a = a_0 + \al &~~~~~~~ b=b_0 + \bt \\ \end{array} \ee
 we find the linearized equations \bea \label{linear_eq}
 \al'' &=& -{6 \over \rho^2}\, \al -{(4+2) \over \rho}\, \al' -{2 \over \rho}\,
 \bt' \non
 \bt'' &=& -{6 \over \rho^2}\, \bt -{(4+2) \over \rho}\, \bt' -{2 \over \rho}\,
 \al' \non
 0 &=& \al'' + {2 \over \rho} \al' + \bt'' + {2 \over \rho} \bt' \eea
Due to the symmetry in exchanging $\al$ and $\bt$ it is easy to
diagonalize by choosing $\al^+=\al + \bt, ~\al^-=\al-\bt$. The
solutions are \be \begin{array}{rclrcl} \label{linear_soln}
 \al^+ &=& c^+_1\, \rho^{s^+_1} + c^+_2\, \rho^{s^+_2} &~~~~~~~ s^{+}_{1,2} &=& -1,\, -6 \\
 \al^- &=& c^-_1\, \rho^{s^-_1} + c^-_2\, \rho^{s^-_2} &~~~~~~~ s^-_{1,2} &=& (-3 \pm i \sqrt{15})/2  \\
  \end{array} \ee
where $c^\pm_{1,2}$ are arbitrary constants. The interesting
modes are the modes of $\al^-$, since the solution with $s^+ =
-6$ is forbidden by the first order constraint, while the $s^+ =
-1$ is simply a translation in $\rho$ (a diffeo) which had to be
there since the equations are $\rho$ independent.

We now proceed to solve the non-linear set of equations
(\ref{def_cone_eq}). Clearly the linear approximation is expected
to hold only for large $\rho$ when the corrections are small. We
are looking for a solution where for $\rho \to 0$ $e^{2\, a}$ will
combine with $d\rho^2$ to make a smooth 3d space, while $e^{2\,
b}$ approaches a finite limit. Therefore we expand around the
singular point $\rho=0$ taking the following ansatz \bea
\label{def_ansatz}
 a(\rho) &=& \log(\rho) + a_1\, \trho^2 + a_2\, \trho^4 + a_3\,
 \trho^6 + a_4\,
 \trho^8 \dots \non
 b(\rho) &=& \log(\rho_0) + b_1\, \trho^2 + b_2\, \trho^4 + b_3\,
 \trho^6 + + b_4\, \trho^8 \dots  \eea
where $\trho := \rho / \rho_0$, and we hope the ansatz is self
consistent and allows to solve for the series coefficients
$a_i,\, b_i$ order by order. Indeed, we find using Mathematica
\cite{Mathematica} {\it the equations to be solvable} and \be
\begin{array}{rclclrclcl}
 a_1 &=& -1/18            &\simeq& -5.6 \cdot 10^{-2}    & ~~~b_1 &=& 1/6             &\simeq&  1.7 \cdot 10^{-1} \\
 a_2 &=& 67/8100          &\simeq&  8.3 \cdot 10^{-3}    &  ~~b_2 &=& -13/540         &\simeq& -2.4 \cdot 10^{-2} \\
 a_3 &=& -4031/2679075    &\simeq& -1.5 \cdot 10^{-3}    &  ~~b_3 &=& 191/42525       &\simeq&  4.5 \cdot 10^{-3} \\
 a_4 &=& 163777/535815000 &\simeq&  3.1 \cdot 10^{-4}    &  ~~b_4 &=& -33391/35721000 &\simeq& -9.3 \cdot 10^{-4} \\
\end{array} \ee

After finding these solutions I was disappointed to find that
their existence was already discussed in the mathematics
literature \cite{Bohm1,Bohm2} where similar Ricci-flat metrics of
cohomogeneity one are described which approach a cone as a limit,
and that the initial value problem was discussed in
\cite{EschenburgWang}, although these papers do not consider going
through the cone to achieve a topology change.

\subsection{The cone and the action}
\label{cone_action}

Let us estimate the dependence of the action on the smoothing
parameter $\rho_0$. Since the metrics are solutions for all $\rho$
the bulk term vanishes and only the boundary term contributes to
the action. We saw that at large distance distances the
corrections to the metric behave as $h \sim (\rho/\rho_0)^{-3/2}$.
Thus $I \sim h^2 \sim \rho_0^3/G_5$. The proportionality factor
could include a $\log(\rho_1/\rho_0)$ factor as well as an
arbitrary multiplicative constant, where $\rho_1$ is a long
distance cut-off on the validity of the local model, $\rho_1$
being of order $L$ which is the only length scale at
$\mu_{\mbox{merger}}$. The multiplicative constant can and in
general will be different on the two sides of the transition,
producing a high derivative ``kink'' in $I$, namely a {\it phase
transition point}.


Despite the success of describing the a Ricci-flat topology change
using the cone over $\stwosq$ it turns out that this {\it cone is
unstable}.\footnote{I thank I. Klebanov for suggesting this, and
M. Rangamani and E. Witten for sharing this unpublished result
\cite{RangamaniWitten}.}

Let us first verify this result and then discuss the implications
for the phase diagram. The ansatz for the perturbation is \be
\label{stab_ansatz}
 ds^2 = d\rho^2 + {\rho^2 \over 3}
 \left( e^{2\, \eps}\, d\Omega^2_{\theta,\phi} +e^{-2\, \eps}\, d\Omega^2_{\chi,\sigma} \right). \ee
Namely, the unstable mode, $\eps(\rho)$, is a change in relative
size of the two spheres. The action for this mode is \be
\label{stab_action} I = \int \left({4\, \pi \over 3}\right)^2\,
\rho^4 \left[ 12(\cosh(\eps)-1)/\rho^2 - \eps'^{~2} \right]. \ee
Since $F=-I/\bt$ we see that $\eps(\rho)$ indeed has a negative
mass-squared (actually, due to the volume factor it is not
evident yet that the mode is tachyonic as seen below is detail).

The instability is alarming at first, but it can merge nicely in
the picture. The point is that if the cone is cut-off both at
small and large $\rho$ then it becomes stable. The cut-off at
small $\rho$ is supplied by $\rho_0$ while at large $\rho$ the
local cone description ceases at a length-scale $\rho_1$ of the
same order as $L$ or the Schwarzschild radius.

Let us estimate when should this instability appear. The
quadratic part of the action, throwing out overall constants is
\be \label{eps_action}
 I \sim \int \rho^4\, d\rho \left[ {6 \over \rho^2}\, \eps^2 -
\eps'^{~2} \right] \ee
 Looking for a marginally stable mode (a zero-mode) we should
solve the equations of motion $\eps'' + 4\, \eps'/\rho + 6\, \eps
/\rho^2 =0$, whose solutions are $\eps(\rho)=\sum_{i=1}^2\, c_i\,
\rho^{s_i}$ where the exponents are given by $s_{1,2}=(-3 \pm i
\sqrt{15})/2$, which are naturally the same exponents that
appeared in the linearized analysis (eq \ref{linear_soln}). Hence
if Dirichlet boundary conditions are put at $\rho_0,\rho_1$
marginal stability will occur at \be
 \rho_1/\rho_0 = \exp(\pi/\mbox{Im}(s)) \simeq 5.1 \ee
 Alternatively, the action (\ref{eps_action}) can be recast in
canonical form \be
 \label{eps_action2}
 I \sim \int d\hrho \left[ {2 \over 3 \, \hrho^2}\, \eps^2 - \eps'^{~2} \right] \ee
 through the change of variables $\hrho^{-1} = 3\, \rho^3$.

\section{Generalization and critical dimension}
\label{generalization}

Let us consider generalizations of the previous section, section
\ref{local}. Instead of starting with a background $\IR^4 \times
\IS^1$ we could have taken $\IR^{d-1} \times \IS^1$ with $d \ge 5$
creating a transition in which the $\IS^{d-3}$ of the string
horizon shrinks, and is replaced by an $\IS^2$ connecting the
north and south pole of the BH along the $z$ axis. This would
lead us to study cones over $\IS^2 \times \IS^{d-3}$.

Here we choose to study more generally cones over $\IS^m \times
\IS^n$ because we can do this more general case for the same
price. It would be interesting to find which of these and of other
general cones can be produced by replacing $\IS^1$ by a more
general compactifying manifold.

It will turn out that all the important properties of these cones
will depend only on the total dimension $d=m+n+1$, and so it is
natural to speculate that this property is more general.

The Ricci flat cone is given by \be \label{cone_gen}
 ds^2 = d\rho^2 + {\rho^2 \over d-2}
 \left( (m-1)\, d\Omega^2_{\IS^m} + (n-1)\, d\Omega^2_{\IS^n} \right), \ee
where the constants are required in order to get the correct
Einstein constant for the base, namely $d-2$ ($\IS^m$ has Einstein
constant $m-1$).

The ansatz for a deformed cone is \be \label{def_cone_ansatz_gen}
 ds^2 = d\rho^2 + e^{2\, a(\rho)}\, d\Omega^2_{\IS^m} + e^{2\, b(\rho)}\,
  d\Omega^2_{\IS^n} \ee
and Einstein's equations are \bea \label{def_cone_eq_gen}
 a'' &=& (m-1)\, e^{-2\, a} -m\, a'^{~2} - n\, a'\, b' \non
 b'' &=& (n-1)\, e^{-2\, b} -n\, b'^{~2} - m\, a'\, b' \non
 0   &=& m\,(a'^{~2} + a'') + n\,(b'^{~2} + b''). \eea
One verifies again that the equations imply a first order
constraint which is consistent with the second order equations,
and reduces the number of required initial conditions to 3.

The cone solution corresponds to
$a_0=\log(\rho)+\log(m-1)/2-\log(d-2)/2,
~b_0=\log(\rho)+\log(n-1)/2-\log(d-2)/2$. Let us linearize first
the equations around this solution. Denoting \be
\begin{array}{lr}
 a = a_0 + \al &~~~~~~~ b=b_0 + \bt \\ \end{array} \ee
 we find the linearized equations \bea \label{linear_eq_gen}
 \al'' &=& -{2\,(d-2) \over \rho^2}\, \al -{2\,m + n \over \rho}\, \al' -{n \over \rho}\,
 \bt' \non
 \bt'' &=& -{2\,(d-2) \over \rho^2}\, \bt -{2\,n + m \over \rho}\, \bt' -{m \over \rho}\,
 \al' \non
 0 &=& m\, (\al'' + {2 \over \rho} \al') + n\, (\bt'' + {2 \over \rho} \bt') \eea
The equations are diagonalized  by choosing $\al^+=m\, \al + n\,
\bt, ~\al^-=\al-\bt$. The solutions are \be
\begin{array}{rclrcl} \label{linear_soln_gen}
 \al^+ &=& c^+_1\, \rho^{s^+_1} + c^+_2\, \rho^{s^+_2}
 &~~~~~~~ s^{+}_{1,2} &=& -1,\, -2\,(d-2) \\
 \al^- &=& c^-_1\, \rho^{s^-_1} + c^-_2\, \rho^{s^-_2}
 &~~~~~~~ s^-_{1,2} &=& {d-2 \over 2}\,(-1 \pm i \sqrt{1-{8 \over d-2}})  \\
  \end{array} \ee
where $c^\pm_{1,2}$ are arbitrary constants. Again the interesting
modes are the modes of $\al^-$, since the solution with $s^+ =
-2(d-2)$ is forbidden by the first order constraint, while the
$s^+ = -1$ is simply a translation in $\rho$ which had to be there
since the equations are $\rho$ independent.

Note that $d=10$ is a critical dimension for the discriminant.
For $d \to \infty$ the two roots are real and approach $-(d-2)$
and $-4$.

Looking for a smoothed cone we repeat the expansion of
$a(\rho),\, b(\rho)$ near $\rho=0$ as in eq (\ref{def_ansatz}),
and find \be \label{def_ansatz_gen} \begin{array}{rclrcl}
 a_1 &=& {\frac{-n\, \left( n -1  \right) }
      {6\,m\,\left( m + 1 \right) }},
 & b_1 &=& {\frac{\left( n - 1 \right) }{2\,\left( m + 1 \right)}} \\
 a_2 &=&
    {\frac{{\left(n-1 \right) }^2\,n\,
        \left( 18\,m^2 - 3\,n + 17\,m\,n \right) }{180\,m^2\,
        {\left(m+1\right) }^2\,\left(m+3\right) }}
 & ~~b_2 &=&
    \frac{- {\left(n-1 \right) }^2\,
          \left( 3 + 3\,m + 2\,n \right) }{12\,
        {\left(m+1 \right) }^2\,\left(m+3 \right) } \\
 a_3 &=& \frac{{\left(n-1 \right) }^3\,n\,
     \left( 405\,m^4 + 30\,n^2 - 119\,m\,n^2 +
       27\,m^3\,\left( 15 + 32\,n \right)  +
       m^2\,n\,\left( 270 + 461\,n \right)  \right) }{5670\,m^3\,
     {\left(m+1\right) }^3\,\left(m+3 \right) \,\left(m+5 \right)
     } & \\
  b_3 &=& \frac{{\left(n-1 \right) }^3\,
     \left( 15\,m^3 + n^2 + m^2\,\left( 45 + 24\,n \right)  +
       m\,\left( 30 + 30\,n + 11\,n^2 \right)  \right) }{90\,m\,
     {\left(m+1 \right) }^3\,\left( 15 + 8\,m + m^2 \right) }
     & \\
 \end{array} \ee

Let us turn now to the stability analysis. The fluctuations we
want to analyze are \be \label{stab_ansatz_gen}
 ds^2 = d\rho^2 + {\rho^2 \over d-2}
 \left( (m-1)\, e^{2\, \eps/m}\, d\Omega^2_{\IS^m}
  +(n-1)\, e^{-2\, \eps/n}\, d\Omega^2_{\IS^n} \right). \ee

The action for this mode is \be \label{stab_action_gen}
 I = \int c\, \rho^{d-1} \left[ {1 \over \rho^2} \left( m\, (d-2)\, e^{-2\, \eps/m} + n\, (d-2)\, e^{2\, \eps/n}
-(d-1)\,(d-2) \right)
 -  ({1 \over n}+{1 \over m})\, \eps'^{~2} \right] \ee
where the constant is given by $c=\Omega_m\, \Omega_n\,
(m-1)^{m/2}\, (n-1)^{n/2}\, (d-2)^{-(d-1)/2}$ and $\Omega_m =
(m+1)\, \pi^{(m+1)/2}\, \Gamma((m+1)/2+1)$ is the volume of
$\IS^m$.
 The quadratic part of the action, disregarding overall
constants is \be \label{eps_action_gen}
 I \sim \int \rho^{d-1}\, d\rho \left[ {2\,(d-2) \over \rho^2}\, \eps^2 -
\eps'^{~2} \right] \ee The zero mode equation is $\eps'' + (d-1)\,
\eps'/\rho + 2\,(d-2)\, \eps/\rho^2 =0$. The solutions are
$\eps(\rho)=\sum_{i=1}^2\, c_i\, \rho^{s_i}$ where the exponents
are given by \be \fbox{$ \room
 s_{1,2}={d-2 \over 2}\,(-1 \pm i
\sqrt{{8 \over d-2}-1}) $}\ee

Again these are the same exponents we got in the linearized
analysis. We see that the field $\eps(\rho)$ is unstable only for
$d<10$ while for $d=10$ it is marginal in the quadratic
approximation. Let us tabulate \be
\begin{array}{c|cccccc}
 d                      &  5   & 6    & 7    & 8    & 9  & 10 \\
 \hline
 \mbox{Im}(s)^2         & 15/4 & 4    & 15/4 & 3    & 7/4  & 0 \\
 \exp(\pi/\mbox{Im}(s)) & 5.06 & 4.81 & 5.06 & 6.13 & 10.7 &
 \infty \\
 \end{array} \ee
 In general \be \fbox{$ \room 4\, \mbox{Im}(s)^2=16-(d-6)^2  $}\ee
  (\cite{BKdraft}, a closely related formula appeared in
 \cite{GibbonsHartnoll} eq. (34)) and
 \be \mbox{Re}(s)= \left\{
 \begin{array}{cr}
 -(d-2)/2 & \mbox{  for  } d \le 10 \\
 -(d-2)/2\, (1 \pm \sqrt{1-{8 \over d-2}}) & \mbox{  for  } d \ge 10
 \end{array} \right. \ee (see figure \ref{res}).

\EPSFIGURE{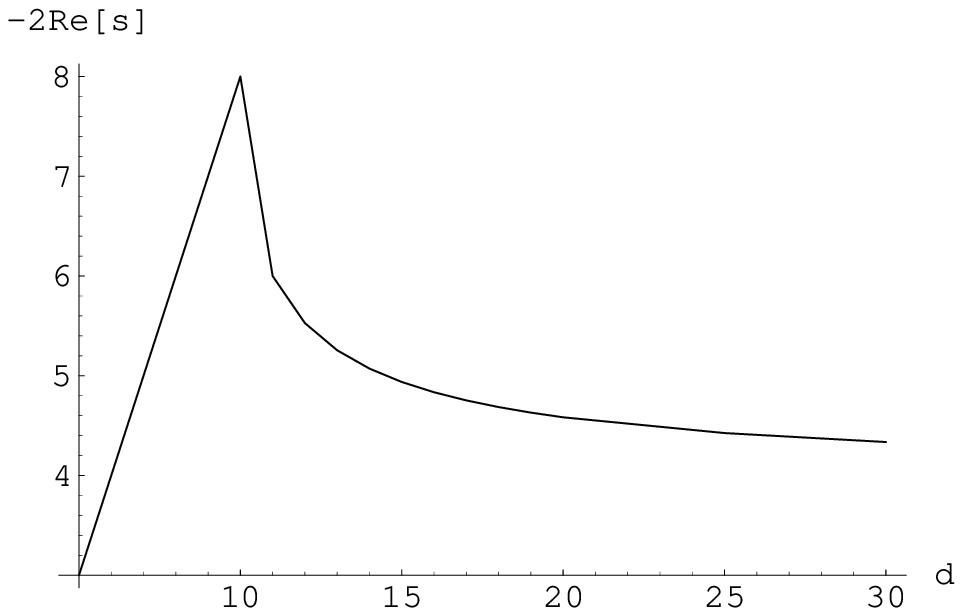}{A plot of $-2\, \mbox{Re}(s_-)$ as a function
of the dimension $d$. \label{res}}

Alternatively, the action (\ref{eps_action_gen}) can be recast in
canonical form \be
 \label{eps_action2_gen}
 I \sim \int d\hrho
 \left[ {2 \over (d-2) \, \hrho^2}\, \eps^2 - \eps'^{~2} \right] \ee
 through the change of variables $\hrho^{-1} = (d-2)\, \rho^{d-2}$.

\section{Summary and Discussion}
\label{Discussion}

\subsection{Summary of suggested phase diagram}
\label{summary_suggested}

Let us recapitulate the suggested picture for the transition which
emerges from our discussion (figure \ref{phasediag2}). The
transition has a typical behaviour for a first order transition,
in the presence of an even free energy, the only stable phases are
the black-hole and the uniform black-string and there is an
unstable non-uniform black-string phase. Starting with the string
and reducing $\mu$ (string $\to$ BH) one reaches first $\mu_1$ the
point of first order transition to the BH phase. However, since
the transition occurs via tunnelling due to statistical and/or
quantum fluctuations the string is still meta-stable. If we
continue to lower $\mu$ we reach the Gregory-Laflamme point,
$\mu_{GL}$, where a marginally tachyonic mode appears and the
system decays immediately. For lower $\mu$ the string is still a
solution, though unstable, and we encounter ``copies'' of the
transition at $\mu \to \mu/k, ~k \in \IN$.

Going in the opposite direction, BH $\to$ string, one starts with
a 5d black-hole at small $\mu$. As $\mu$ is increased one
encounters the first order point $\mu_1$ beyond which the BH is
only meta-stable and later one reaches a perturbative instability.
Altogether the system traces a ``hysteresis'' curve typical of a
first order transition.

In our first suggestion the perturbative instability was thought
to occur at $\mu_{\merger}$ which is where the black hole phase
terminates. However, it turns out that the cone over $\stwosq$ is
unstable (\cite{RangamaniWitten} and related results in
\cite{Duff:1984sv} for $AdS_4 \times M_p \times M_{7-p}$,
\cite{BerkoozRey} for $AdS_7 \times S^2 \times S^2$ and
\cite{HerzogKlebanov} for $AdS^5 \times S^3 \times S^2$), and a
dramatic dependence on the dimension is found in which cones over
products of spheres have an instability for $d<10$ which
disappears for $d>10$. Hence for $d>10$ I view the original
suggestion (figure \ref{phasediag2}) to be viable, and a small
feature must be added, namely a high derivative kink (2rd
derivative kink in 5d) to the curve $\mu=\mu(\lambda)$ at the
merger point (see subsection \ref{cone_action}). For $5 \le d <10$
on the other hand the picture must be more interesting and we must
amend the phase diagram at the vicinity of the merger transition
expecting the perturbative instability to occur before merger,
roughly when $r_1/r_0 \simeq D \simeq 5$ where $r_1$ represents
the radius of the spheres at the large length cutoff of the cone -
at the outer edge of the cone, and $r_0$ is the radius of the
minimal $\IS^2_{\theta,\phi}$.

It is tempting to speculate further about the form of the phase
diagram in the vicinity of the merger transition. Consider
decreasing $r_0$ while $r_1$ stays roughly constant. The resolved
cone continues to be a solution as we saw, but each time we cross
a threshold $r_1/r_0 \simeq D^k, ~k \in \IN$ a new marginally
tachyonic mode emerges. The situation is just like moving along
the string phase - each time a critical point is reached two
phases emerge from this point. A detailed local analysis is
required in order to determine whether these new phases are stable
or not (the analog of the analysis of \cite{Gubser} at the GL
point). One possibility is that the phase diagram near the merger
point will look like repetitious copies - see figure
\ref{phase_insert}(a). Since the cone is self-similar and the new
phases appear again and again in a self similar way it is tempting
to speculate about a more involved possibility with a {\it
fractal} phase diagram and {\it fractal} physics near the merger
point (see figure \ref{phase_insert}(b)), in which the constant
$D$ will be related to the fractal dimension. This will be in tune
with the numerical discovery of scaling behavior in General
Relativity \cite{Choptuik_scaling}.

\EPSFIGURE{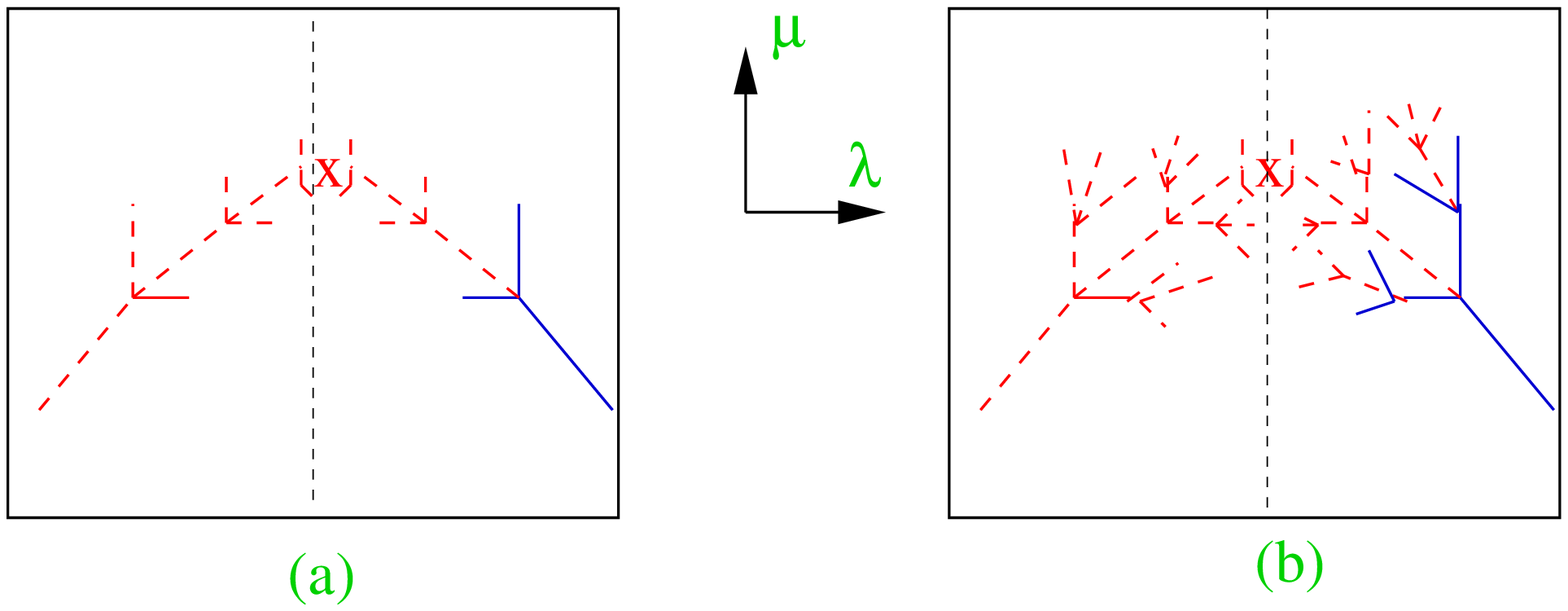,width=12.5cm}{Possible behaviors around
the merger point for $5 \le d <10$. (a) minimal change. (b) A
fractal diagram. The thin dashed line separated the black hole
phase on the right from the unstable string to the left. The
number of negative modes for each phase is suppressed here.
\label{phase_insert}}

The topology changing transition does not occur for $d<5$.
Physically it is because that would require analyzing black holes
in 3d and lower which do not have regular horizons but rather
induce deficit angles, while mathematically it is because both
sphere factors in $\IS^m \times \IS^n$ must have positive Einstein
constant and so $m,\,n \ge 2$. For $5 \le d < 10$ the picture
stays similar to $d=5$.

However, for $d > 10$ the previously tachyonic function stabilizes
due to a competition in the action between the volume factor and a
certain mass-like term. Closely related dependence on the
dimension was observed in geometric studies of minimal cones
\cite{Hsiang_cones}; in construction of cohomegeneity one Ricci
flat metrics \cite{Bohm1,Bohm2};  in the stability of AdS
compactifications over a product space \cite{DFGHM} and
\cite{Shiromizu:2001sp} (this result could be related to this one
by the cone-AdS relation \cite{KlebanovWitten}, though here D
brane probes do not exist as such); in \cite{BKdraft} and in the
stability of Schwarzschild black holes in a asymptotically conical
background \cite{GibbonsHartnoll}. However, here they arise in a
natural dynamical context for the first time. I can see no
connection between the 10d criticality observed here and the
critical dimension of superstring theory, and so if such a
connection existed it would be very interesting.

A phenomenon that could remind one of the Gross-Witten transition
as claimed in \cite{SusskindGW} was found, namely, that the free
energy probably has a kink on the wall in configuration space
where a topology change takes place (subsection
\ref{cone_action}). In 5d the jump is in the 3rd derivative (or
higher), and in other dimensions it is at least $-2\,
\mbox{Re}(s_-)$ (figure \ref{res}). Moreover, a high order phase
transition from black hole to non-uniform black string seems to
occur at the merger point, but this is not a real phase
transition, unless the emanating non-uniform black string is
stable (which I do not expect). This issue should be revisited
when the phase diagram in the vicinity of merger transition is
better understood.

\subsection{No Cusps}

The {\it cusps} on the horizons which were discovered in numerical
simulations of black hole collisions (see \cite{Lehner_review}
and references therein) {\it were not seen} in the local model
which has smooth minimal spheres. That seems to indicate that the
cusps are velocity dependent, vanishing for a static
configuration. It would be interesting to test this by measuring
numerically the cusp for various relative velocities keeping equal
separations. Another possibility is that the local model is not
sensitive enough to detect the cusps.

\subsection{Topology change}

The term ``topology change'' in General Relativity is being used
with several different meanings, for instance the topology of the
whole space-time could change as a parameter is varied or the
topology of a spatial slice could change as time progresses. The
topology change here will be of the first kind, evolving with a
parameter, just like the flop \cite{flop} and conifold
\cite{conifold} transitions for 6d Calabi-Yau surfaces. Expected
implications for time evolution are briefly discussed below, and
we should note that classical no-go theorems \cite{Geroch,Tipler}
are not valid due to the presence of the singularity (somewhat in
the flavor of the possible classical topology changes of
\cite{Horowitz_change}).

Let us compare the merger transition discussed here with the
conifold transition \cite{conifold} and the ``$G_2$-conifold''
\cite{GPP_G2,AtiyahWitten}. The merger transition is in $m+n+1$
dimensions and centers around the cone over $\IS^m \times \IS^n$
(minimum 5d) while the conifold ($G_2$-conifold) is in 6d (7d)
involving the cone over $\IS^3 \times \IS^2$ ($\IS^3 \times
\IS^3$). Despite the similar topologies and Ricci flatness of the
geometries they differ in supersymmetry and metric: whereas the
previous topology changes had Killing spinors the merger cone does
not, and indeed it is not even stable. Accordingly the metric on
the spheres differs - the conifold ($G_2$-conifold) metric is
given by the homogeneous space $T_{1,1}=(SU(2) \times SU(2))/U(1)$
($SU(2)^3/SU(2)_{\mbox{diag}}$) whereas the merger cone involves
the round spheres and in 6d can be written as $T_{1,0}=SU(2)/U(1)
\times SU(2)$.

\subsection{Future directions}

I would like to point out the following topics for future work
\begin{itemize}
\item Phase diagram near the merger point.
\item Lorentzian signature
\item Explosive transition
\item Uniqueness theorems (no hair)
\item $\al '$ corrections
\item Merger and quantum gravity
\end{itemize}
which will be discussed below one by one.

The suggested phase diagram (figure \ref{phasediag1}) was found
to require corrections near the merger point at least for $5 \le
d <10$. Some possibilities were discussed above in subsection
\ref{summary_suggested}.

So far we worked only in the Euclidean signature, and it is an
important question to understand the Lorentzian. The static
solutions can be readily turned to static Lorentzian solutions
outside the horizon, and one expects an extension into the inside
of the horizon to exist. However, the singular cone solution at
the merger transition will translate into a {\it naked
singularity} on the horizon, at which point the structure of the
light cone may be different and a problem with the extension of
the solution could arise. Even more interesting is the {\it time
dependent} behavior, when the tachyon decays. For instance, one
could start with an unstable string $\mu < \mu_{GL}$ with initial
conditions that as $t \to -\infty$, both the tachyon its time
derivative approach zero $ ~T,\dot{T} \to 0$. Since the problem
is time dependent the apparent horizon does not coincide anymore
with the event horizon, and it would be interesting to understand
the Penrose diagram.

The hysteresis curve contains two sections of tachyon decay, one
starting after the Gregory-Laflamme point, and the other close to
the merger transition. Since the energy emitted during these
decays scales as the mass of the black object, it must equal a
numerical constant times the mass (there are no dimensionless
parameters at the transition points). Assuming large extra
dimensions one finds large fusion/fission explosions
\cite{explosive}.

The black hole uniqueness theorems (``no hair''), which
characterizes a time-independent 4d black hole by its mass and
angular momentum (and possibly gauge charges), does not hold as
such for $d>4$. The 5d rotating black ring solution of
\cite{EmparanReall} is a beautiful counter-example where for some
values of the mass and angular momentum 3 solutions exist - 2
black rings and 1 black hole. The black-hole black-string
transition provides another example for non-uniqueness above 5d,
since at some values of $\mu$ there are at least 3 solutions - a
(perturbatively) stable string, an unstable non-uniform string and
the black hole. This example is static, but the asymptotic
space-time is not Euclidean (though it is flat). It would be a
shame to loose such a nice principle at higher dimension, and it
would be better to try and ``fix'' it. Following this work a
natural speculation arises - that the topology of the horizon must
be specified, and that unstable solutions should not count
\cite{uniqueness}.

It would be interesting to find string theoretic $\al'$
corrections to the cone geometry. On the one hand large
corrections could appear close to the tip where curvatures
approach the string scale, but on the other hand the cone
topology seems to be essential for the topology change, which may
protect the cone. Perhaps non geometric modes become important
like in the conifold transition \cite{StromingerConifold}.

Finally it is natural to wonder about the quantum gravity
interpretation of a black hole merger, as mentioned in the
introduction, and it is hoped that this work makes a step in that
direction.

\vspace{0.5cm} \noindent {\bf Acknowledgements}

\noindent
 It is a pleasure to thank Lenny Susskind for introducing
me to this problem some years ago;
 Gary Horowitz for rekindling my interest and for discussions;
 John Bahcall for encouragement;
 Tsvi Piran and Evgeny Sorkin for collaboration on a related project and comments on the paper;
 Sunny Itzhaki and Juan Maldacena for some enjoyable discussions;
 Steve Gubser, Igor Klebanov, Mukund Rangamani, Edward Witten for correspondence and important comments;
 David Berenstein, Sergey Cherkis, Eric Gimon and other members of the IAS group for discussions;
 and finally the University of California at Santa Barbara
 and Stanford university for hospitality while this work was in gestation.

The first version of this paper appeared while I was at the School
of Natural Sciences, Institute for Advanced Study, Princeton and
the work was supported by DOE under grant no. DE-FG02-90ER40542,
and by a Raymond and Beverly Sackler Fellowship. Currently my work
is supported in part by The Israel Science Foundation (grant no
228/02) and by the Binational Science Foundation BSF-2002160.


\appendix
\section{Comments on a local analysis around the GL critical
point}

In this appendix I would like to describe an approach to the
determination of the local properties of the system at the GL
point which is somewhat different then the one used in
\cite{Gubser} and is actually standard in the thermodynamics of
phase transitions. It requires the computation of solutions up to
the second order in perturbation rather than third, but the
required hard computations were not pursued to the end.

Start by considering the action functional, namely working in the
canonical ensemble. At the GL point the quadratic potential has
one zero mode, and the objective is to go beyond the quadratic
approximation, and determine whether this mode is negative or not,
which are the only qualitative possibilities. This will determine
the current goal, namely, whether the emerging phase is stable or
not, and whether its $\mu$ starts climbing down or up.

The quadratic approximation to the action can be written as \be
\label{I2}
 -I_2[x,y_i;\hmu]= k_0\, \hmu\, |x|^2 + \sum_i m^2_i |y_i|^{~2}
 \ee
 where $x$ is the marginally stable mode, $y_i$ are all the other
modes diagonalized such that their mass-squared are $m^2_i$,
$\hmu=\mu-\mu_{GL}$ and  $k_0$ is a positive constant. One would
like to trace the emerging solutions $x,\, y_i,\, \hmu$ as a
function of a perturbation parameter $\eps$. Clearly at first
order only $x$ can be turned on and the first task is to find it
explicitly by solving the equations of motion for $I_2$.

Due to the $U(1)_z$ isometry of the uniform black-string all modes
with $z$-dependence are complex with a specific charge $q_i$ such
that under a rotation of the $z$ axis by an angle $\theta$ they
transform by \be y_i \to e^{q_i\, \theta}\, y_i \ee By the
construction of $x$ we have $q_x=1$. Let us use the fact that the
action is real and invariant under this action. The first
consequence is that cubic terms involving only $x,\bar{x}$ (which
would imply a first order transition) are forbidden. Therefore we
must go to quartic order in $\eps$. For this we must expand the
action to include \be
 -I_4=  \sum_i g_i\, y_i\, x^2 + \mbox{c.c.} + \sum_j h_j\, y_j\, |x|^2  + k\, |x|^4 \ee
where the sum is over modes of charge $\pm 2$ or 0 which couple
to quadratics in $x,\bx$ and $g_i,\, h_j$ are the respective
couplings.

One takes \bea x=x^{(1)}\, \eps + x^{(2)}\, \eps^2 \non
     y_i = y_i^{(2)}\, \eps^2 \eea
And finds that $-(I_2 + I_4)$ becomes a quadratic form in
$\eps^2$, and its signature is determined by the sign of \be
 k - \sum_i |g_i|^2/m^2_i - \sum_j h_j^{~2}/m^2_j \ee

In practice it is more convenient to find the solution for
$x,y_i$ up to second order in $\eps$ and check the sign of the
resulting action $-(I_2 + I_4)$ - a positive mode means that it
is a minimum and hence a second order phase transition will form,
whereas a negative mode means that there is already another
minimum lower than the present one, and so the system goes
through a first order transition (see figure \ref{2tran}).

If one computes the constant $k_0$ in $I_2$ (\ref{I2}) one can
solve for $\eps=\eps(\hmu)$ and extract quantitative information
about the initial rates of change of $I,\, x,\, y_i$ along the new
phase as a function of $\mu$.


\end{document}